# Ferroelectric-like SrTiO$_3$ surface dipoles probed by graphene

Ray Sachs, Zhisheng Lin, and Jing Shi*

Department of Physics and Astronomy, University of California, Riverside, CA 92521

The electrical transport properties of graphene are greatly influenced by its environment. Owing to its high dielectric constant, strontium titanate (STO) is expected to suppress the long-range charged impurity scattering and consequently to enhance the mobility. However, the absence of such enhancement has caused some controversies regarding the scattering mechanism. In graphene devices transferred from SiO$_2$ to STO using a newly developed technique, we observe a moderate mobility enhancement near the Dirac point, which is the point of charge neutrality achieved by adjusting the Fermi level. While bulk STO is not known as a ferroelectric material, its surface was previously reported to exhibit an outward displacement of oxygen atoms and ferroelectric-like dipole moment. When placed on STO, graphene shows strong and asymmetric hysteresis in resistivity, which is consistent with the dipole picture associated with the oxygen displacement. The hysteretic response of the surface dipole moment diminishes the polarizability, therefore weakens the ability of STO to screen the Coulomb potential of the impurities.

*Correspondence to: jing.shi@ucr.edu

Numerous experiments [1-14] have been conducted to probe the nature of scattering in graphene such as both long- and short-ranged Coulomb scattering [15, 16], resonant scattering [17], and phonon scattering [18], etc. For example, the effects of the long-ranged charged impurity scattering have been studied by intentionally doping graphene with randomly distributed impurities as potassium [15] and gold [19]. In addition to the confirmation of the linear dependence of the Coulomb scattering conductivity of graphene on carrier density, these studies have also found the decrease in mobility with increased doping concentration, shifting of the Dirac point position with a change in charged state, and an overall independence of the value of the minimum conductivity $\sigma_{min}$ with increased doping, all predicted by the theory [20, 21]. The Coulomb potential of the charged impurities can be modified by introducing screening. The effects of screening have been studied by covering pristine graphene with various dielectric liquids [7-10, 22] and placing graphene on high dielectric constant (high-κ) substrates [6, 7, 11, 23].

Strontium titanate (STO) is an interesting dielectric substrate to use due to the higher $\kappa$. At room temperature, $\kappa_{STO}$ (~200-300) is nearly two orders of magnitude higher than $\kappa_{SiO2}$ (~3.9). At cryogenic temperatures, $\kappa_{STO}$ was reported to increase to 5,000-10,000 [13]. The increased $\kappa$ is expected to suppress the charged impurity scattering and lead to enhanced mobility in graphene with further drastic enhancement expected at low temperature. Cuoto et al. successfully exfoliated and patterned graphene devices onto the surface of 500 μm thick STO (111) [13]. Although the carrier density increased by an order of magnitude from 50 K to 250 mK and the increase in capacitance is evident in the smaller $V_g$ range needed to sweep through the Dirac point at lowering temperature, the mobility remains unchanged. This is apparent in the curves of resistivity $R_{xx}$ vs. carrier density $n$ that overlap for all temperatures. Their explanation is the dominance of short-ranged, resonant scattering which provides a good fit at intermediate carrier densities. A recent theoretical study [14] provides an alternative explanation. A competition of various scattering mechanisms can explain the data with Coulomb scattering necessarily dominant at low density because it is asymptotically the largest at vanishing density. An increase in substrate charged impurities at low temperature was assumed in order to explain the unchanged mobility even with the greatly enhanced screening. An objective of our work is to first carefully examine the state of the STO substrates, both 200 μm thick STO with different orientations and the atomically flat, 250 nm thick epitaxial STO grown on Nb-doped STO (100) substrate prepared by pulsed laser deposition (PLD). In all cases, the transport properties of the *same* graphene sheet are directly compared between $SiO_2$ and STO substrates. The effect of screening does occur as is evident in the graphene mobility enhancement near the Dirac point. Surprisingly, however, the STO surface behaves like a ferroelectric material revealed by the hysteresis in graphene resistivity. Furthermore, the surface dipole moments undergo ultra-slow relaxation when subject to an external electric field.

**Results**

Single-layer graphene has been mechanically exfoliated from Kish graphite onto the surface of doped Si with 300 nm of thermally grown $SiO_2$. The graphene flakes can be easily located via optical microscopy, electrodes are patterned into a Hall bar geometry with electron beam lithography, and metal (Pd or Au) is deposited for good electrical contact.

Raman spectroscopy and electron transport measurements are made while on the surface of $SiO_2$ before it is then transferred to the surface of 200 μm thick STO (MTI Corporation) or 250 nm epitaxial STO. To prepare the surface of the 200 μm thick STO (100), we follow the standard recipe [24], i.e. thoroughly cleaning in ultrasonic baths of acetone then deionized water, and dipping in buffered HF etchant. Some STO (100) substrates are annealed before transfer, but we find that annealing does not make any significant difference to the device characteristics. The Nb-doped STO is prepared in the same fashion prior to PLD deposition of epitaxial STO. After STO deposition, the film is annealed in $O_2$ at 1100 °C for 6 hours, which leads to an atomically flat surface verified by atomic force microscopy. The graphene device is then measured again so a direct comparison of the electronic properties can be made. The device transfer method is outlined in Fig. 1a and is a variation of previous work [25]. The key difference is that an entire pre-fabricated graphene device with metal electrodes can be placed on any arbitrary substrate so that no nano-lithography is needed after it is transferred to the target substrate. In fact, as shown in Fig. 1b, stand-alone single-layer graphene flakes on STO or any other target substrates may be very difficult or even impossible to locate for making contacts with standard nano-lithography. After transfer, the metal electrodes may become detached from the graphene flake, in which case it results in an open circuit and the transfer process fails. Although it is difficult to inspect visually, the success of the transfer process can be easily scrutinized with electrical measurements. We have been achieving an increasing yield with our devices (currently >80%). The results reported in this paper represent over a half dozen of transferred devices.

It is reasonable to consider the degrading effects of the additional PMMA layers on top of the graphene flakes needed for transferring and the NaOH wet etchant. These could add impurities or cause damages to the graphene to increase the Coulomb and other types of scattering. A control experiment was performed to transfer a graphene device from the surface of $SiO_2$ to a different $SiO_2$ wafer. The mobilities before and after transfer are comparable and can either be slightly increased or decreased. This is not surprising due to the changed local environment in the amorphous $SiO_2$ surface such as charge traps and other defects. Similar mobility variations are often seen in non-transferred graphene devices on $SiO_2$. We believe that the transfer procedures themselves do not create any significant change to carrier mobility. Full details of the experiment are the subject of another paper of ours. To further confirm the results in transferred graphene/STO, we have performed another control experiment with the same fabrication method used in ref. 13, and obtained consistent results with the transferred devices.

A peculiar hysteresis effect has been previously observed in graphene devices placed on high-$\kappa$ substrates [7, 13]. This effect is cited as the reason for a small gate voltage $V_g$ range in ref. 13. In our devices, minor hysteresis loops are always present even with small gate voltage ranges. With intermediate gate voltage ranges, we observe strong hysteresis in both resistivity $\rho_{xx}$ and Hall coefficient (see Fig. 2) at room temperature. In comparison, the resistance of the same graphene on $SiO_2$ before transfer shows no hysteresis. In separate devices transferred to $SiO_2$ as our references, we do not observe such hysteresis either. Moreover, in graphene on STO, we do not find any qualitative difference after it is annealed in vacuum at 250 °C for 10 hours. From these facts, we rule out that the transfer process itself is the cause of the observed hysteresis. The positions of the Dirac points are offset between the up and down sweeps in gate voltage. In the down sweep, the Dirac point is located on the negative side; whereas in the up sweep, it is located on the positive side. As a reference, the non-hysteretic data of the pre-transfer graphene/SiO$_2$ are shown in both figures. Another noticeable feature in the figures is the asymmetry in both resistivity and Hall data between the far left and far right. Clearly, at high +$V_g$ where $\rho_{xx}$ approaches the saturation, the saturated value is consistently higher than that on the far left. Fig. 2b indicates that the saturated electron density is lower than the saturated hole density. This asymmetry is observed in all transferred graphene/STO devices.

The strong hysteresis that we observe in all samples is highly dependent on the sweeping range of the gate voltage (see Fig. 3). The separation between the two Dirac points is widened as the sweeping range increases. At the largest range, i.e. from +200 to -200 V, the separation is as large as ~150 V! The asymmetry between the left and right plateaus is more evident. This strong hysteresis in resistivity corresponds well to the carrier density hysteresis loop (Fig. 3b) calculated from the simultaneously measured Hall coefficient $R_{xy}$ using $n = \frac{B}{R_{xy} \cdot e}$ where $B$ is the applied perpendicular magnetic field. In Fig. 3b, the anomalies near the Dirac points are removed and replaced by the interpolation of the data from both sides (See SI). At the zero gate voltage, the residual carrier density is ~ 6x10$^{12}$ cm$^{-2}$, indicating the same amount of the bound charge density on the surface of STO. These loops are reminiscent of the minor loops of ferroelectric materials. Here the hysteresis is not a known property of graphene itself, so it must come from STO.

Indeed, if the conductivity is plotted as a function of the measured $n$ from the Hall data, the hysteresis disappears (see Fig. 3c). This means that once $n$ is set, although it is not a unique function of $V_g$, the conductivity (or resistivity) is uniquely determined. In the pre-transfer state, i.e. graphene/SiO$_2$, the conductivity shows linear dependence on $n$, which is consistent with the charged impurity scattering mechanism. After it is transferred to STO, a sub-linear behavior emerges. The slope slightly increases as the Dirac point is approached from both sides (Fig. 3c), which suggests enhanced mobility. Fig. 4a is the calculated mobility from the conductivity and Hall coefficient for both before- and after-transfer devices. The absolute carrier mobility is somewhat different after transfer, although the two devices are made of the same graphene flake. This is not surprising since the local environment of SiO$_2$ and STO is always different. However, what is more interesting is the marked contrast in the density dependence of the carrier mobility. Let us ignore the region in the vicinity of the Dirac point (represented by open symbols) of both devices where the calculated density deviates from the linear gate voltage dependence which is affected by residual electron and hole fluctuations. In graphene/SiO$_2$, the mobility is essentially constant on the hole side (not large enough gate voltage range on the electron side). On the contrary, the mobility of graphene/STO clearly varies with the carrier density. In the high carrier density limit, the mobility slowly approaches saturation on both electron and hole sides. The smaller the carrier density, the greater the mobility enhancement is. At 4.2 K, the mobility has a greater enhancement. On the hole side, the largest mobility is enhanced by ~ a factor of four. The temperature dependence is distinctly different from the previous result reported in similar graphene/STO device [13]. We argue that this density dependent behavior is consistent with the charged impurity scattering model. At high carrier densities, graphene itself has abundant carriers to screen the nearby charged impurities; therefore, the presence of the substrate does not appear to affect the mobility significantly. However, near the Dirac point where the carrier density is low, the screening ability of graphene itself is greatly limited; consequently, screening from STO plays an important role in reducing impurity scattering. This is very different in $SiO_2$ that is not able to screen the impurities as effectively. The same density-dependent mobility enhancement phenomenon is also observed in graphene on epitaxially grown, 250 nm thick STO (Fig. 4c). Here the thin and atomically flat STO (see SI) was grown by PLD on Nb-doped STO (100) substrate which serves as the back gate. As shown in Fig. 4c, the maximum hole density is $\sim 2\times10^{13}$ cm$^{-2}$, but is achieved with a -4 V gate voltage. This effective gating is due to the much larger capacitance with the thin STO. The qualitatively same mobility enhancement near the Dirac point is observed (Fig. 4c). However, we do not observe a steep increase in mobility at low temperatures in both types of STO. We believe that STO does not behave as a linear dielectric medium. As such, the dielectric constant is not a well-defined quantity. This is different from the response of STO in ref. 13, where STO was kept in the linear response regime by intentionally restricting the gate voltage range. In non-linear media, the screening ability is related to the slope of $n$ vs. $V_g$. Our data indicate that the slope does not change dramatically as the temperature is lowered from 300 to 4.2 K (Fig. 4b), which is in contrast to the response of STO in the linear regime (ref. 13).

The hysteretic behavior is clearly not an intrinsic property of graphene, but of STO. In addition to the hysteresis, the asymmetry in carrier density saturation between the large positive and negative gate voltages is yet another property of STO. Bulk STO is not known to have any ferroelectric transition. After all measurements of the graphene device are finished, we deposit a 100 nm thick Au layer on the entire top surface of the 200 μm thick STO that covers the graphene and the metal contacts. In this parallel-plate capacitor, the STO dielectric material is primarily sandwiched between the two metals and the graphene/STO direct contact only takes a negligibly small portion of the entire area. We measure the ac response of the impedance and find no measurable hysteresis at all in the capacitance of this metal/STO/graphene/metal structure; therefore, the bulk STO does not show any ferroelectricity. However, our observation in graphene/STO devices seems in an apparent contradiction. We attribute the ferroelectric-like behavior in graphene/STO to the surface dipoles on STO that are not significantly modified by the presence of graphene. It was reported in an electron diffraction study that STO has a few atomic layer thick surface in which the oxygen atoms are displaced outward [26]. This is known as "puckering" that results in an inward dipole moment (Fig. 5d). The same phenomenon was observed in both different terminations of the same STO orientation and different STO orientations [27]. This surface dipole layer could be altered by a metal layer deposited on top, in order to explain the absence of any hysteresis in metal/STO/metal capacitors. Nevertheless, the oxygen displacement of the pristine STO is apparently unaltered by graphene; therefore, the dipole moment can be probed by graphene. The microscopic details of the surface atomic arrangement require more sophisticated surface probes to reveal, but it may occur because graphene does not have any mobile atoms as in metals to modify the STO surface energy to reduce the puckering. As a result, the inward dipole moment is responsible for the bias in carrier density, i.e. positive carrier background in graphene that shifts the $n$ vs. $V_g$ hysteresis loop upward (Fig. 3b), which in turns results in the observed asymmetry in the transport properties. As $V_g$ is swept, the displacement of the oxygen atoms may change, which consequently causes the bound charge density in STO to change. This surface layer response to $V_g$ may be significantly different from that of the majority of atoms in bulk STO, and the hysteresis is easily picked up by graphene. Apparently, like any other ferroelectric media, the surface dipole moment does not only depend on the magnitude of $V_g$, but also the history of the $V_g$ sweeps.

**Discussion**

Graphene resistivity is a very sensitive probe to examine both the ground state and the dynamics of the surface dipoles of STO under an external electric field. Fig. 5a shows the sweeping rate dependence of graphene resistivity. A gate voltage is ramped while a determined amount of time lapses before the data point is collected to allow the STO to "relax". By increasing the wait time, the positions of the dual Dirac points shift. At the shortest wait time of 3 minutes, the down sweep peak is on the right at positive $V_g$ and the up sweep peak is on the left at negative $V_g$. Increasing to 6 minutes, the peaks change sides. Increasing to 12 minutes and they are further apart. Such rate dependence does not exist in the resistivity of the pre-transfer graphene/SiO$_2$ or in the capacitance of metal/STO/metal. The rate dependence reveals an ultra-slow relaxation process of the dipole moment of graphene/STO. Another type of relaxation measurement is to set $V_g$ to a given value and record the resistivity as a function of time. The relaxation process strongly depends on the set $V_g$. To compare the relaxation for different set $V_g$'s, we must prepare the identical initial state. We do so by first setting to +200 V and waiting until the resistivity settles. This initial preparation takes between 45 minutes to 1.5 hours depending on the previous $V_g$. Then we set $V_g$ to a final voltage smaller than +200 V in one step and start to record the data as a function of time. As shown in Fig. 5b, if the final $V_g$ is greater than +50 V, the initial relaxation takes ~ two hours followed by even slower relaxation. The same time-dependent process can be tracked by carrier density data measured simultaneously through $R_{xy}$ as shown in Fig. 5c. In principle, the carrier density at t=0 is determined by two parts: instantaneous response due to the abrupt external $V_g$ change and slow relaxation due to the STO surface dipoles. For example, at +200 V, the electron density is ~ $6\times10^{12}$ cm$^{-2}$. After the gate voltage is ramped down to +50 V, the electron density changes very little and remains close to $6\times10^{12}$ cm$^{-2}$ at t=0. This is because the instantaneous electron density decrease is only $\frac{\varepsilon_0}{d}(\Delta V_g)$ ~ $4\times10^9$ cm$^{-2}$. Therefore, most electrons at +200 V are retained by the slowly relaxing surface dipoles. Here we adopt the surface puckering model (Fig. 5d) to explain the transport behaviors. At +200 V, the oxygen displacement is the smallest due to the opposing external electric field. As the final $V_g$ is set greater than +50 V, the oxygen outward displacement increases initially which leads to a decrease in electron density. After a few hours, the slower increase in electron density indicates a very slow inward movement of oxygen atoms toward the STO bulk, but the orientation of the dipole moments remains the same. If $V_g$ is set to or below +50 V, the resistivity increases, reaches a peak, and then decreases. As indicated by the density data in Fig. 5c, the electron density decreases first, crosses zero and then the carrier type switches to holes. The hole density continues to increase over time, which indicates a very slow outward movement of oxygen atoms. This slow adjustment in charge density at both large and small final $V_g$'s is consistent with the Coulomb interaction between the charged graphene sheet and the surface dipoles. Similar slow relaxation has been observed in STO properties [27] and the actual microscopic mechanism was related to the migration of defects in STO [28]. +50 V appears to be the crossover point at which the effects of the surface dipoles and the external electric field from the positive $V_g$ nearly cancel each other. Since small positive electric fields (for $V_g$ < 200 V) do not destabilize the inward dipole moment associated with the oxygen displacement, the surface dipole moment remains in the same direction but can change in magnitude.

The strong hysteresis and ultra-slow relaxation are also observed in STO with (110) and (111) orientations. Similar oxygen displacement induced dipole moments were previously predicted in different orientations [29]. These results indicate that the ferroelectric-like surface dipoles always exist in STO regardless of their terminations and orientations. We note that most studies were done with the STO surface exposed to vacuum which is different from the present study. The non-zero surface dipole moment in graphene/STO itself suggests that the puckered surface state is a robust ground state of STO due to its own unique electronic structure.

In summary, when graphene is placed on flat STO surface, the screening effect modifies the scattering potential and consequently the carrier mobility of graphene near the Dirac point where the graphene intrinsic screening ability is weak. A surface such as STO which exhibits hysteresis is not optimal for obtaining high carrier mobility in graphene. With our transfer technique, a graphene device can be placed on any surface without the need for locating for nano-fabrication. This opens up possibilities of obtaining high graphene mobility with high dielectric constant substrates. Furthermore, graphene does not appear to quench the surface dipole moment of STO and is found to be very sensitive to the state of the surface dipoles. If combined with more elaborate surface science techniques such as low-energy electron diffraction (LEED) and transmission electron microscopy (TEM) as well as first-principles calculations, graphene can potentially serve as a unique and convenient probe to reveal the microscopic structure and dynamic properties of STO and other oxide surfaces.

**Acknowledgements**

RS and ZSL are supported by NSF/NEB. JS is supported by DOE. We thank Z.Y. Wang, P. Odenthal, H. Wen, R. Kawakami, J. H. Chen, J.D. Zhang, and V. Aji for their technical assistance and useful discussions.

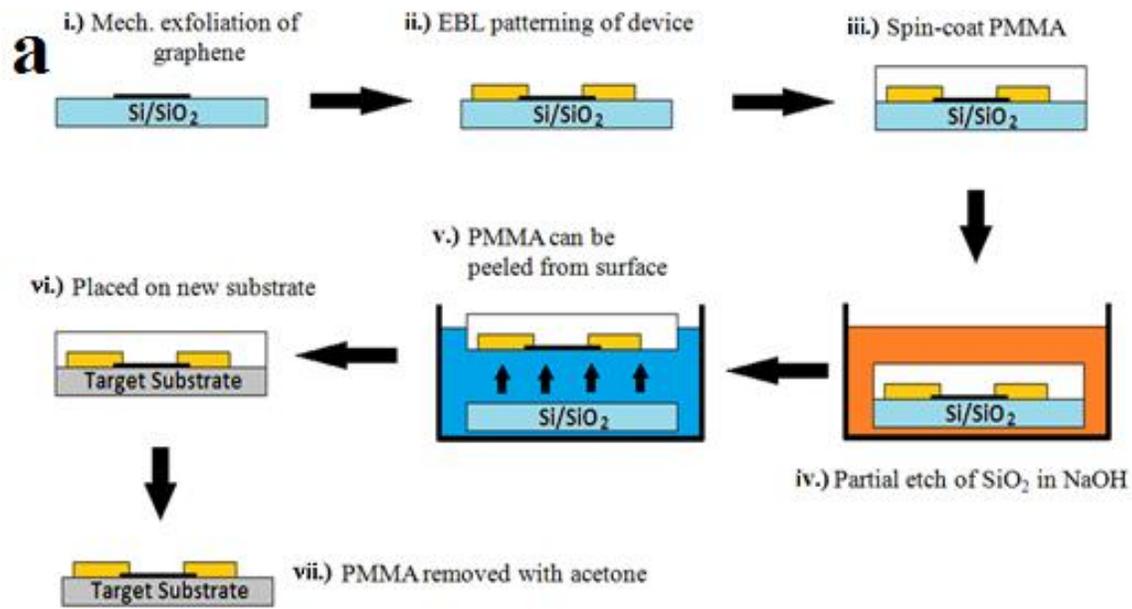
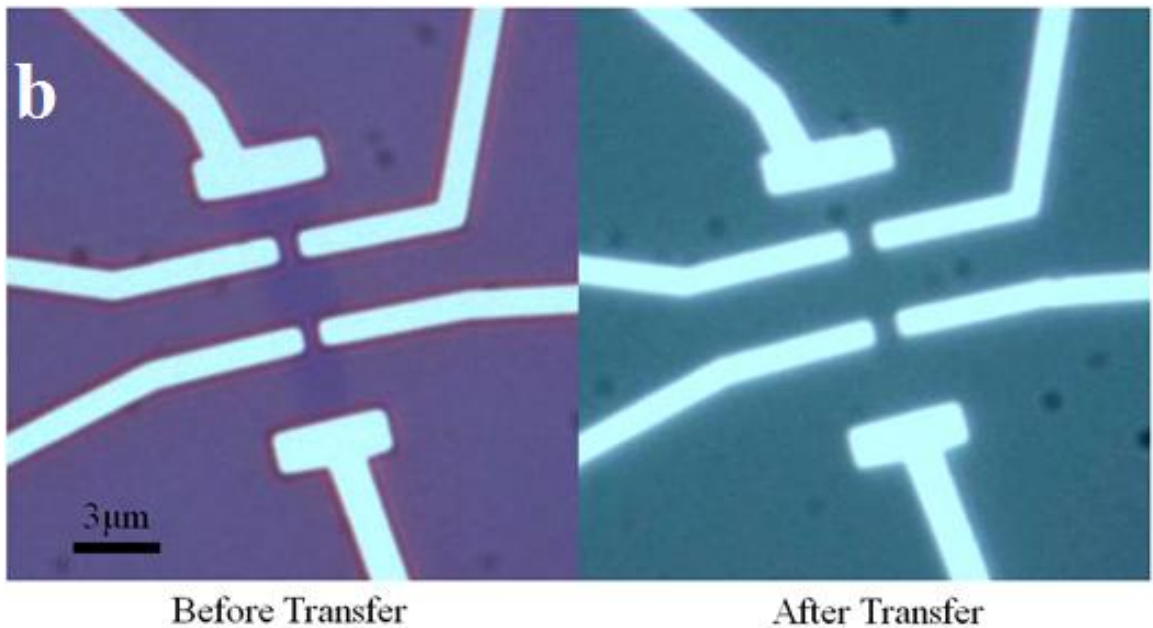

Figure 1. Device transfer technique (a) (see SI). Optical images of graphene device on SiO$_2$ and after transfer on 200 μm STO (b).

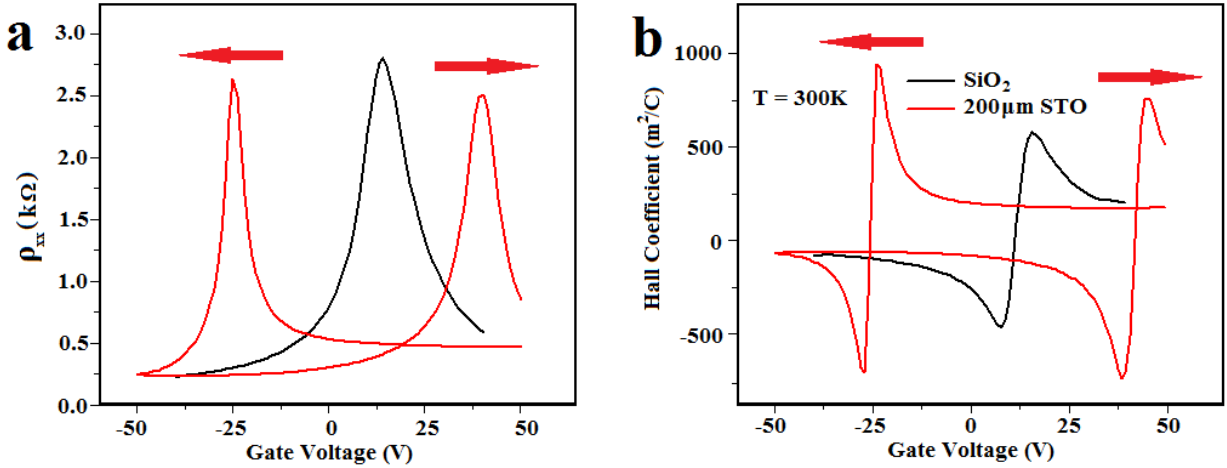

Figure 2. Sheet resistance (a) and Hall coefficient (b) vs. $V_g$ for graphene on $SiO_2$ (black) and 200 μm STO (red) at room temperature. Arrows indicate direction of $V_g$ sweeping.

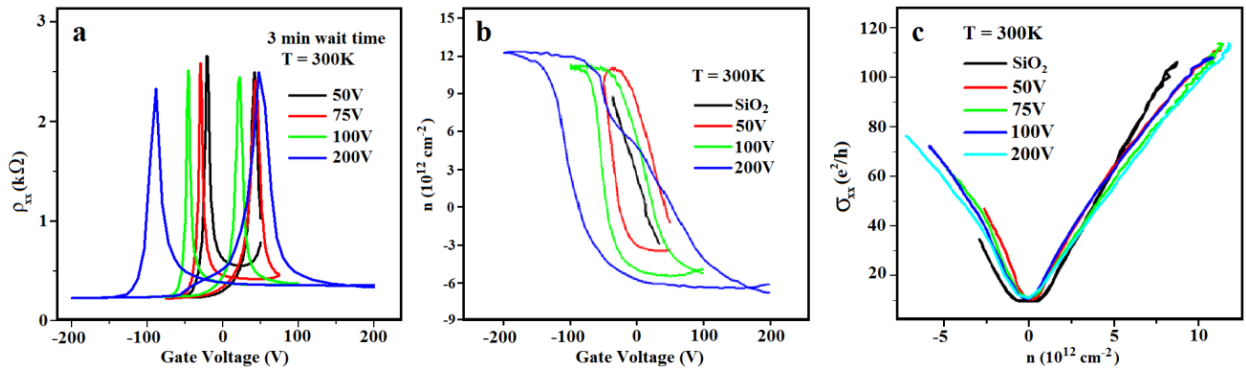

Figure 3. Back gate sweep with varying $V_g$ ranges (a). $n$ vs. $V_g$ showing ferroelectric-like hysteresis (b). $\sigma_{xx}$ vs. $n$ exhibits no hysteresis (c).

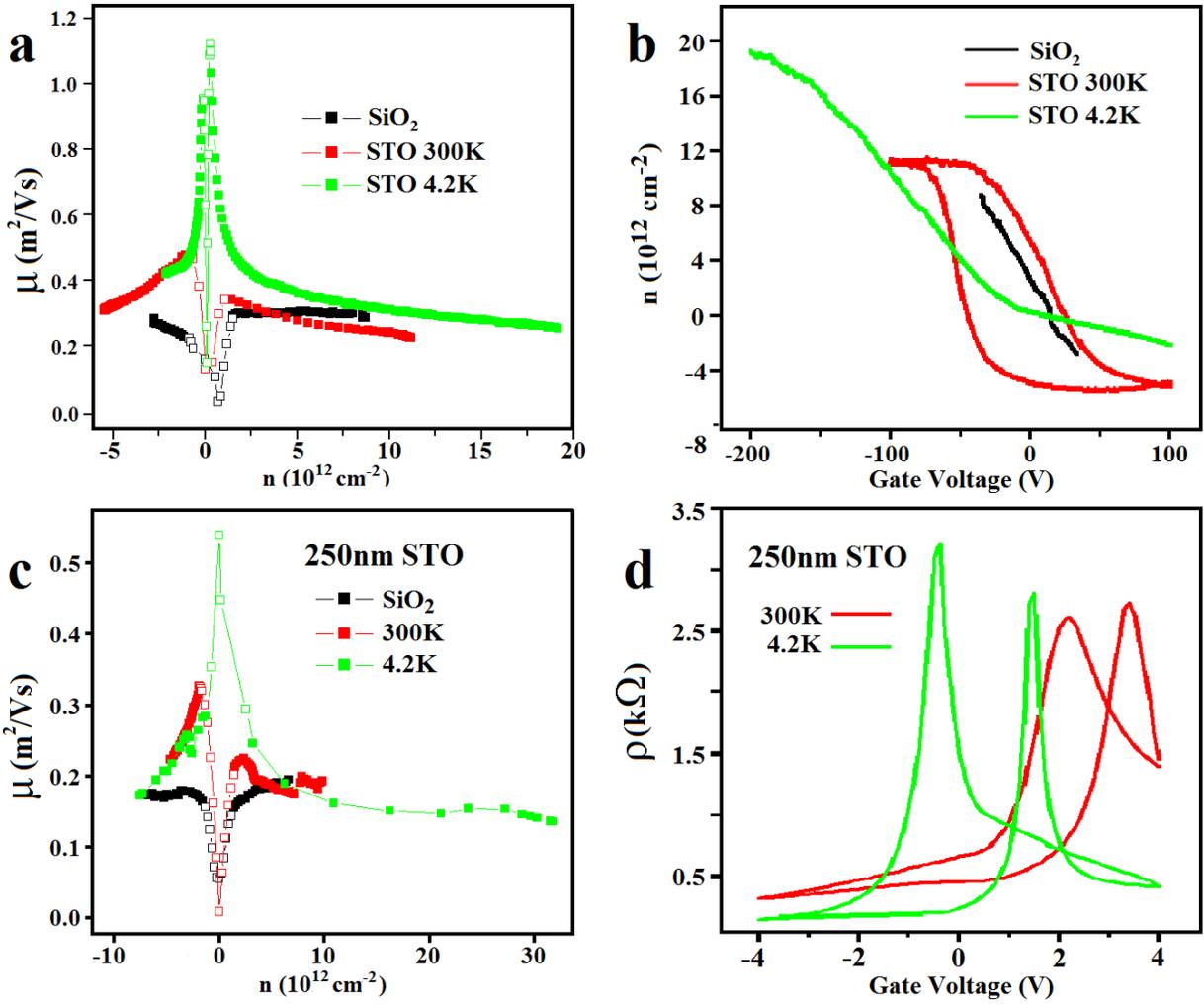

Figure 4. $\mu$ vs. $n$ showing enhancement in Dirac region with increased κ (a). $n$ vs. $V_g$ of graphene on SiO$_2$ and STO at high and low temp (b). Graphene on 250 nm epitaxial STO shows enhanced $\mu$ in Dirac region (c) and also has improved capacitance as evident from the smaller $V_g$ range needed to sweep the resistivity through Dirac point (d).

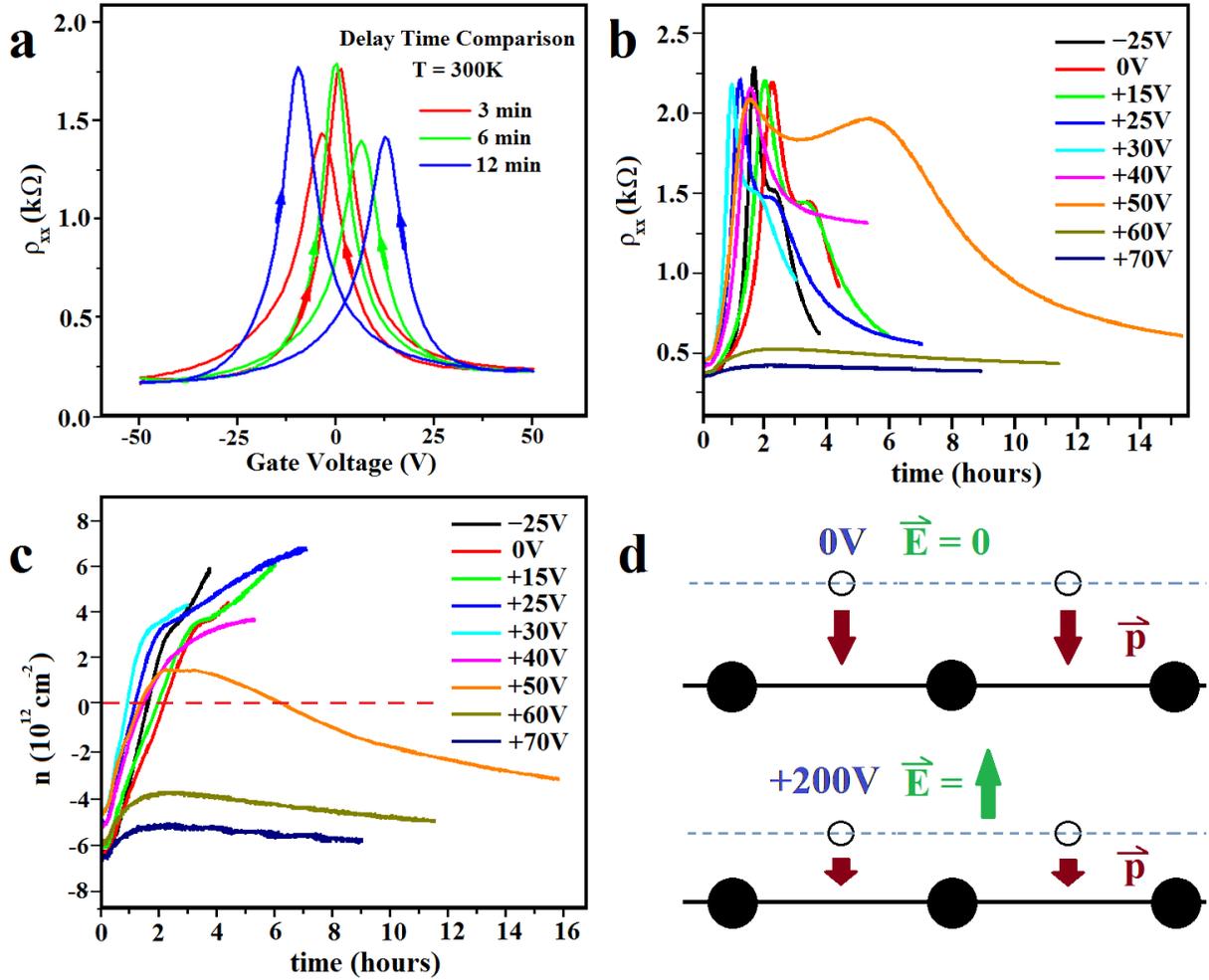

Figure 5. $\rho_{xx}$ vs. $V_g$ with varying delay time (a). Dual Dirac peaks switch position with increasing delay. $\rho_{xx}$ vs. time (b) and $n$ vs. time (c) showing ultra-slow relaxation of surface dipoles. All measurements begin at stable state of +200 V. Carrier type switches when ramped below +50 V with no change at values above. At +50 V the carrier type switches twice. Surface dipole model (d) showing oxygen atoms (open circles) puckered out from the surface at zero applied field. At +200 V, the applied $E$- field causes oxygen to move toward the Sr or Ti atoms (black circles). This decreases the dipole moment.